\documentclass[sigconf]{acmart}
\usepackage[utf8]{inputenc}
\usepackage{graphicx} 
\usepackage{pgfplots}
\usepackage{multirow}
\usepackage{tabularx}
\usepackage{array}
\usepackage{siunitx}
\usepackage{listings}
\usepackage[figuresright]{rotating}
\newcolumntype{?}{!{\vrule width 1pt}}
\usepackage{caption}
\usepackage{subcaption}

\copyrightyear{2021}
\acmYear{2021}
\setcopyright{rightsretained}
\acmConference[K-CAP '21]{Proceedings of the 11th Knowledge Capture Conference}{December 2--3, 2021}{Virtual Event, USA}
\acmBooktitle{Proceedings of the 11th Knowledge Capture Conference (K-CAP '21), December 2--3, 2021, Virtual Event, USA}
\acmPrice{}
\acmDOI{10.1145/3460210.3493556}
\acmISBN{978-1-4503-8457-5/21/12}

\begin{document}
\fancyhead{}

\title{Order Matters: Matching Multiple Knowledge Graphs}

\author{Sven Hertling}
\email{sven@informatik.uni-mannheim.de}
\orcid{0000-0003-0333-5888}
\affiliation{%
  \institution{Data and Web Science Group}
  \city{University of Mannheim}
  \country{Germany}
}

\author{Heiko Paulheim}
\email{heiko@informatik.uni-mannheim.de}
\orcid{0000-0003-4386-8195}
\affiliation{%
  \institution{Data and Web Science Group}
  \city{University of Mannheim}
  \country{Germany}
}

\begin{abstract}

Knowledge graphs (KGs) provide information in machine interpretable form.
In cases where multiple KGs are used in the same system, that information needs to be integrated.
This is usually done by automated matching systems. 
Most of those systems consider only 1:1 (binary) matching tasks. Thus, matching a larger number of knowledge graphs with such systems would lead to quadratic efforts.
In this paper, we empirically analyze different approaches to reduce the task of multi-source matching to a linear number of executions of binary matching systems.
We show that the matching order of KGs and the multi-source strategy actually matter and that near-optimal results can be achieved with linear efforts.

\end{abstract}

\ccsdesc[500]{Information systems~Data extraction and integration}
\ccsdesc[300]{Information systems~Information extraction}

\keywords{Knowledge Graph; Multi-source matching; Incremental merge}

\maketitle

\section{Introduction}
\label{sec:intro}

During the past years more and more knowledge graphs (KGs) and ontologies were generated. The Linked Open Data (LOD) Cloud\footnote{\url{https://lod-cloud.net}} counts more than 1,250 datasets \cite{schmachtenberg2014adoption}, and the LOD-a-lot dataset \cite{lodAlot} has more than 650k datasets. 
Querying and utilizing this data in a uniform way is only possible if the schema and instances are aligned because each dataset can use their own identifiers (URIs) for the same concept.
One key idea of the Semantic Web is to connect representations of the same entities with the \texttt{owl:sameAs} relation. There are also dedicated properties for specifying that two URIs describe the same class (\texttt{owl:equivalentClass}) and property (\texttt{owl:equivalentProperty}).

Usually, these knowledge graphs contain too many resources (classes, properties, and instances) to be manually mapped, and thus many matching systems were developed in the past.
Starting from 2004, the Ontology Alignment Evaluation Initiative \cite{euzenat2011ontology} (OAEI) takes care of evaluating such matching systems. The matchers are uploaded by developers and executed by track organizers.
The matching interface originates from \cite{euzenat2007} and consists of one function which expects two KGs and an optional input alignment which contains already aligned entities.
The result of this function is an alignment which consists of correspondences.
Since 2018, the OAEI has a dedicated track for matching knowledge graphs \cite{dbkwikResults}.

Similarly, many systems are developed in the field of entity resolution/alignment. The main difference to KG matching is the static schema which is required by most systems. In other words, those systems expect the schemas of the KGs to be already aligned.

Due to the interface of all matching systems, they can only align two KGs at the same time, but the usual problem at hand is to align multiple data sources together.
Even the conference track \cite{oaeiconference} at OAEI is originally a multi-source matching problem which is broken down into 21 binary test cases.
If the ultimate goal is to obtain an alignment between $n$ sources, this would require $O(n^2)$ executions of a matching system.

The goal of this paper is thus to reuse the well researched and developed 1:1 matching systems for multi-source matching.
The canonical approach matches each source pair, but it has a quadratic runtime.
As an alternative, we propose an incremental approach, which matches and merges KGs successively.
We show that such incremental merge based approaches can reach near-optimal results in linear runtime.
Moreover, we analyze different ordering schemes for the incremental approaches, and analyze their impact on the matching result.

The contributions of this paper are:
 \begin{itemize}
    \item An extension of the general matcher and evaluation interface in MELT \footnote{\url{https://github.com/dwslab/melt}} (an open-source matching framework recommended by OAEI),
 	\item an implementation of different multi-source approaches in the aforementioned framework,
 	\item an evaluation of those approaches, and
 	\item analysis of the impact of different orderings and multi-source approaches.
 \end{itemize}

The paper is structured as follows: Section \ref{sec:related} discusses the related work about multi-source matching and cluster repair strategies.
It is followed by the description of the different approaches and orderings (Section \ref{sec:approach}). In the subsequent sections, we outline our experimental setup, discuss results, and conclude with an outlook on future works.

Throughout the paper, we assume that the knowledge graphs to be matched are duplicate free in themselves (which is the case most of the time). In case they are not, the sources could be deduplicated \cite{elmagarmid2006duplicate} before running the matcher. This is also in line with most of the matching systems, which enforce a 1:1 alignment as an output, i.e., they assume that each concept has at most one correspondence. Furthermore, we only consider alignments using the equivalence relation, and ignore other possible types of alignment relations, such as \texttt{subclass} or \texttt{part of}.
Even though there is no official definition of a \emph{knowledge graph} \cite{KGRefine}, in this paper, we use the term to denote an RDF based dataset with an explicit ontology (terminology box) and instances (assertion box).
This also requires that the matchers produce correspondences on both levels.

\section{Related Work}
\label{sec:related}
Starting in 2018, the OAEI campaign \cite{oaei2018} has a new knowledge graph track which requires instance and schema matches.
Over the years, developers adapted their matching components such that they return useful results.
Therefore, systems like AgreementMakerLight (AML) \cite{aml2020}, LogMap~\cite{logmap2020}, ATBox~\cite{atbox2020}, and ALOD\-2Vec~\cite{alod2vec2020} are able to match the schema as well as the instances.
Another related approach is PARIS \cite{suchanek2011paris} which computes probabilities for each correspondence at the relation, instance, and schema level and uses this information to improve and validate other correspondences. Therefore, the instance alignment can help to improve the class alignment and also the other way around.
However, all those systems are designed in a fashion that they match two knowledge graphs at a time and do not provide means for multi-source matching.

In the field of entity resolution and record linkage, there are some approaches which support the case of multiple sources \cite{famer,survey}.
Thus, the closest related work to the presented one here, is multi-source entity resolution which covers a similar matching task, but does neither align the schema nor reuse it to improve performance.
One reason is that in a typical entity resolution scenario, all sources have to have the same schema (all instances are from the same class and have the same -- or at least manually matched -- properties). 

Also in that field, \emph{blocking} \cite{blocking,blockingsurvey} is usually used as a strategy for restricting the search space. Based on a rather simple condition like sharing the same label, all resources are clustered. Each resource in a cluster will thus have the same label and possibly many of them represent the same real world concept.
This matching strategy is most common in the field of entity matching \cite{wang2011entity}. As shown in \cite{linkdiscoverysurvey} the blocking key for instances can be the corresponding classes.
Here, the same observation holds: blocking keys can only be defined if the schemas of the data sources are already matched. 
Moreover, blocking keys usually need to be defined per class. In large, cross-domain knowledge graphs, there are often thousands of classes \cite{heist2020knowledge}, all of which would require the definition of different blocking keys. 

An overview of integrating multiple sources is given by Rahm~\cite{surveyrahm}.
In this work, the integration of KGs is also discussed as one use case among others, such as integrating open data. It is stated that the main challenges are data quality and semantic heterogeneity~\cite{dong2014knowledge}.
The presented core idea of the overview paper is to do holistic data integration which starts with one source as an initial KG. Then, further KGs are merged to it. For each entity, it has to be decided if it should be merged to an existing entity or if it is a new concept not appearing yet. The order of how this is done, plays an important role. However, to the best of our knowledge, there is no work so far which analyzes the impact of the ordering.

The holistic clustering approach mentioned before is implemented by~\cite{nentwig2017distributed} in a distributed manner for Linked Data. It uses Apache Flink for parallelization and clusters each entity by simple 
linguistic similarity on labels. Afterwards, the type of the entity is used to further refine the clusters.
Another parallel implementation from the same department is called \texttt{Dedoop}~\cite{kolb2012dedoop} which does the entity resolution in a map reduce fashion. 

The difference is still that the schema is considered to be the same or at least matched.
However, for popular KGs, such as DBpedia~\cite{auer2007dbpedia}, YAGO~\cite{yago}, or Wikidata~\cite{vrandevcic2014wikidata}, this assumption cannot be made.
\cite{haas2009schema} argued that not only the data, but also the schema needs to be integrated. They also propose a holistic approach based on rules, but only examples are given and no system was implemented nor evaluated.

When multiple sources are integrated, a candidate pre-selection is usually the first step. Afterwards, the similarities within each cluster are computed. Based on this similarity graph,
various cleaning and improvement strategies can be applied. The main idea behind such algorithms is to use the graph structure and confidence values to find strongly connected components which usually represent one concept.
If multiple resources are weakly connected, this can be a signal for wrongly stated correspondences.
One approach was implemented by Raad et al.~\cite{raad2018detecting} which computes an error score for each \texttt{owl:sameAs} link in the LOD cloud.
It is based on the community detection algorithm Louvain~\cite{blondel2008fast}. 
The FAMER system~\cite{famer} also has an incremental clustering and repairing step after the linking phase. Multiple clustering approaches are compared such as \texttt{Center}, \texttt{Star}, and \texttt{CLIP}.

Summarizing, while systems for knowledge graph matching exist, they do not consider the case of multi-source matching. On the other hand, multi-source entity matchers assume a pre-aligned schema, which is not given for most real-world knowledge graphs, or only look at entity matching. 
Hence, to the best of our knowledge, this paper is the first to address both of these challenges simultaneously.
Moreover, for multi-source matching, the order in which the sources are processed is assumed to have an impact on the results, but that impact has not been analyzed so far.

\section{Approach}
\label{sec:approach}

In the following section, different approaches are described which solve the task of multi-source matching reusing 1:1 matching systems.
In comparison to the usual matching interface \cite{euzenat2007}, the new interface consists of a function which expects a list of KGs, an input alignment, and additional parameters. The return value is again an alignment which is represented as a set of correspondences of the form $\langle entityOne, entityTwo, =, 1.0 \rangle$. Throughout the paper, this result is also called system alignment. Even though multiple sets of same entities can be used to represent the alignment, we stick to the usual format to be able to capture similarity graphs (a graph where the edges represents correspondences).

\begin{figure}
	\centering
	\includegraphics[width=\linewidth]{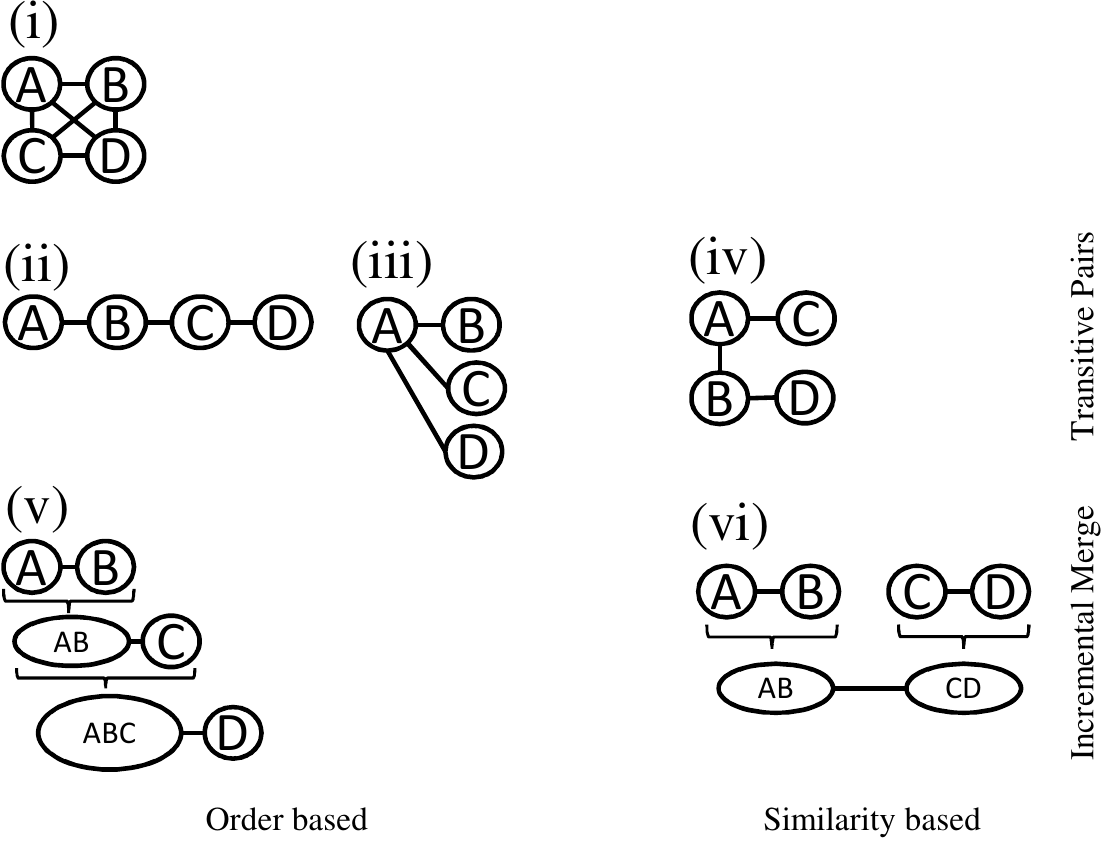}
	\caption{Strategies to reduce a multi source matching problem to 1:1 matches. The second row contains approaches called \texttt{transitive pairs} whereas the third row covers the \texttt{incremental merge} strategies.
		(ii), (iii), and (v) need an order of KGs to compute the final execution plan. (iv) and (vi) are both based on textual similarity of the given KGs.}
	\label{fig:matching}
\end{figure}

Figure \ref{fig:matching} shows an example of four KGs A, B, C, and D, which are matched with different multi-source approaches.
All of them will be explained in more detail in the following sections.

\paragraph{(i) All Pairs}
In this approach, all possible combinations of the KGs are matched.
It is assumed that the order of source and target of a 1:1 matching problem is irrelevant. This means the result of matching source A to source B is the same as matching B to A.
Therefore, it is a combination without repetition taken two KGs at a time.
The number of executions is ${n\choose 2}=\frac{n*(n-1)}{2}$ where $n$ is the number of source KGs, i.e., it is quadratic in the number of knowledge graphs (unlike the other approaches discussed below). The runtime is the sum of time used to match each combination.

\subsection{Transitive Pairs} 
Approaches with this strategy use a linear set of $n-1$ matching problems. Matches between other KGs are then found by creating the transitive closure of the resulting alignments. All of those approaches require an ordering of KGs. In this paper, multiple orderings are tried out. The orders are defined by extracting different KG measures such as the number of classes, instances, or the total number of statements (model size). It might sometimes be the case that multiple of these measures produce the same order. All measures can be used in ascending and descending order which doubles the number of possible orderings.

\paragraph{(ii) Transitive Pairs: Windowing}

The goal of this approach is to reduce the number of matching executions while keeping at least a (transitive) connection between all sources.
It should be clear that these approaches reduce the number of executions with the drawback that not all matches might be found. E.g. if there is a concept in source A named $A_i$ and corresponding match in source C named $C_i$, then it is possible to find the correspondence $<A_i, C_i, =, 1.0>$ only in the case where source B also contains a resource $B_i$ which should be matched both $A_i$ and $C_i$.
Creating the execution plan for this approach requires an order of the KGs.
Here the windowing approach is followed, which means that the first KG (in the example it is A) is matched to the second KG (B) and then the second KG to the third one and so on (window size of two).

\paragraph{(iii) Transitive Pairs: First vs Rest}

This approach is similar to the previous one but it will not match the first one (e.g. the one with the highest or lowest number of classes) with the second one in the list.
Instead the first KG is chosen as a hub and all other KGs are matched to it. The complexity and runtime is thus the same as in the previous approach.

\paragraph{(iv) Transitive Pairs: Similarity}

Approaches (ii) and (iii) need an ordering to obtain the final execution plan.
It is also clear that if a concept is not appearing in all KGs, that ordering can lead to missed correspondences.
E.g. if source $C$ and source $D$ in approach (iii) contain a similar concept which is not contained in source $A$, that correspondence cannot be found.
In the general case of KG matching, all of these sources can have different main topics or cover a variety of domains.
This can also be observed in the Linked Open Data Cloud which lists multiple datasets clustered to domains like geography, government, life sciences, media etc.

The main idea of similarity based approaches is to find KGs with a similar topic and to match them first.
Therefore, a minimum spanning tree of these sources are computed with Kruskal's algorithm \cite{kruskal}.
Computing such a tree always requires weighted edges which should represent in this case the topical closeness of KGs.

In this paper, we use textual signals to compute the topical similarity of two KGs.
The weighting is computed by the cosine similarity of tf-idf vectors to mitigate the impact of terms which often appears (such as upper level ontology terms). Those vectors are generated by iterating over all literals in the KG which contains text\footnote{Literals which either have a language tag or the datatype \texttt{xsd:string}, or \texttt{rdf:langString}} and all URI fragments (part after the last slash ``/''  or hashtag ``\#'').
Literal texts are processed by sentence splitting, tokenization, lowercasing, stopword removal, and stemming. The fragments are tokenized additionally with camel case\footnote{\url{https://en.wikipedia.org/wiki/Camel\_case}} and some common separators such as minus (-), underscore (\_), and tilde ({\raise.17ex\hbox{$\scriptstyle\sim$}}). The sentence splitting is left out for the fragments because it usually does not form a whole sentence.

Computing the final execution plan works as follows:
(1) For each source KG, the tf-idf vector based on the texts in the KG is computed. (2) The cosine similarity for each possible matching pair is calculated.
(3) The distances are sorted in ascending order. (4) Each pair is added as long as there is no cycle / the source and target KG does not belong to the same transitive closure.
For an efficient implementation of the last step, we implemented a disjoint-set data structure\footnote{\url{https://en.wikipedia.org/wiki/Disjoint-set\_data\_structure}} to create the transitive closure of KGs and to check if a cycle would be generated.
It is based on the algorithm described in \cite{transitiveclosure} (Section 3.3).

In this approach, it is quite likely that hubs emerge. In Figure \ref{fig:matching}, assuming that A and C are from one domain and B and D are from another, a new KG E from the first domain would be matched to B and another KG F from the second domain would be matched to A such that KGs A and B both become hubs.

\subsection{Incremental Merge} 
Incremental Merge based approaches, in contrast to the transitive pairs ones, do not leave the input knowledge graphs untouched, but merge two KGs after matching them. This might add some computational complexity, but the assumption is that the merged KGs then contain more information and can be matched better to other KGs. Moreover, the problem of missing single correspondences, as discussed above, can be mitigated in those cases.

\paragraph{(v) Incremental Merge: Order Based}
The order based approach matches the first two KGs together, creates the union of them, and continues with the next KG in the list.
All available orderings from (ii) can be reused.

One important step is to create the union of the two KGs given the alignment between them.
This merging is performed as follows: First, we determine which of the two KGs acts as a source and which as target.
A merged KG has always precedence for being the target. In case both KGs are leaf nodes, then the one with more triples is selected as the target.
Because no assumption about the storage of KGs is made, leaf nodes which first acts as a target during the merge, are copied. This prevents modifications of KGs which are stored on disk (e.g. Jena TDB or HDT file \cite{HDT}).
With such a storage option, also huge KGs can be processed.

Now that the source and target are fixed for the merge, all triples of the source are merged into the target KG.
All subjects, predicates, and objects of each triple are checked and resources (represented by a URI) are replaced with the URI of the target concept. 
The potential disadvantage of such a merge strategy is that all information/triples of a resource are added to the union. This means every wrong match of two concepts, also adds e.g. the label to the merged concept.
It will only happen when the system is able to cope with multiple labels or other multi-valued properties.
In the end, this can lead to topic/concept drifts which are most often analyzed in the time dimension \cite{topicDrift,topicDriftTwo}.

Preventing such a concept drift is done by adding only information/triples of non-matched entities to the union.
Such a merging strategy has to be defined on a level of RDF triples. If the subject of a source triple has a mapping to a concept in the target KG, this triple does not need to be added to the union
because it will only provide more information to an already existent concept. If only the predicate is mapped (due to the fact that also the schema is matched), the information is new and can be added to the union.
If a resource in the object position has a correspondent concept, this is again an information about an already existent concept in the union. Therefore, it is not added. To sum up this kind of strategy, only triples where the subject or object is not matched, will be added to the union. This prevents a topic drift because the first seen concept is fixed and not further modifications are made even when similar concepts appear in other sources.

After one merge is finished, the result is used directly in the next run of the binary matcher. Thus, systems which are aware of being used in such a scenario, can reuse a potentially build index for the next run. For such a warm start, all triples which are added to the union, are also fed into the matching system for updating necessary indices (i.e., it does not reload the knowledge graphs from scratch in the next run).
If the matcher has no warm start feature, in each run a larger KG needs to be read.

The complexity of approach (v) is $n-1$ but the runtime is calculated by the sum of time used by the matcher plus the merge time after each run.

\paragraph{(vi) Incremental Merge: Similarity}

Similar to the approach (iv), the incremental merge strategy cannot only be used with the order based approach but also with the textual similarity of KGs.
The idea is again to first match sources which share the same topic and then all others. The actual execution plan is calculated by a hierarchical agglomerative clustering \cite{hierarchicalClustering}.
The result is a hierarchy of clusters which can directly be used as an execution plan. In the beginning, each KG is its own cluster and the algorithm searches for the next best KG to form a new cluster.
The similarity between clusters can be determined with single, average or complete link strategies \cite{hierarchicalClustering}. The clustering input is the tf-idf vector of each KG which is described in (iv).
In comparison to the order based incremental search, it is now possible that two already merged KGs should be matched together. In such a case the larger union in terms of statements is selected as the target except that one of them is the result of a previous matching operation. The reason is that the matcher with the warm start feature needs to only update its index with the given subset of triples.

\subsection{Other approaches}

There are more approaches on how to reduce the multi-source KG matching task to multiple binary matching problems which are not evaluated in this paper.
An example of such an approach would be to create the union of all KGs and match the union to itself.
Such approaches are not included because they will not respect the goal of a 1:1 matching system to provide at most one corresponding concept for a given resource.
Thus, most matching systems will only return the correspondences which match the element to itself because it will be the best match.
Another approach of this kind would be to match source A to the union of all KGs except A and repeat this for each source. This would solve the before mentioned issue but again it is not clear to which concept of the union it is matched because the union can contain multiple relevant concepts (from difference sources).

\section{Evaluation}

During the evaluation the datasets and matching systems from OAEI are used.
Many advantages result from this setting: (1) the matching systems are available and tested on the given datasets, (2) the gold standards evolved over time and are freely accessible, and (3) the datasets cover ontology matching as well as instance and schema matching datasets. The evaluation was executed on a server with 768GB of RAM and 40 CPUs (2.4 GHz) with Debian 10 operating system and Openjdk version 1.8.0\_162.

\subsection{Datasets}
The conference dataset \cite{conferenceDataset} is selected as a representative of a multi-source ontology matching problem. It consists of 16 ontologies which cover the domain of conference organization.
Although the initial problem is to combine many ontologies, the actual test cases are reduced to a 1:1 matching problem with a corresponding gold standard.
Out of these 16 ontologies, only seven are involved in the reference alignment which results in 21 test cases with a gold standard. More information on the dataset can be found at the official web page\footnote{\url{http://oaei.ontologymatching.org/2020/conference/index.html}} and in \cite{conferenceDataset}.

The next track is called Knowledge Graph\footnote{\url{http://oaei.ontologymatching.org/2020/knowledgegraph/index.html}} and requires schema as well as instance matches. It consists of eight KGs which cover different topics from the entertainment and gaming domain. They are generated by running a modified version of the DBpedia extraction framework \cite{dbpedia} on Wikis from the Fandom\footnote{\url{https://www.fandom.com}} Wiki hosting platform \cite{dbkwik}. 
All in all, three main topics exist: Star Wars, Marvel, and Star Trek. For five pairs of sources, a reference alignment exist.
Together, they contain 49 class, 181 property, and 15,129 instance correspondences \cite{dbkwikResults}. The latter ones are extracted directly from the Wiki pages by inspecting links in sections with a header containing the word \textit{link}. Nearly 25 \% of these matches are non-trivial (meaning they cannot be found by simple string comparison). The schema alignments are generated by experts. The whole gold standard in this track is partial because not all resources in the KG are matched.

\subsection{Systems}
In our experiments, we select the top performing systems from OAEI 2020 \cite{oaei2020} in the respective tracks.
This results in the following matching systems: AML, LogMap, ALOD2Vec, and ATBox.
Due to the fact that LogMap runs into deadlocks in the KG track, it is replaced by the PARIS matching system \cite{suchanek2011paris}.
The latter one is specifically developed for matching the schema as well as the instances and is therefore a perfect replacement.
The equivalence result files of the last iteration of PARIS are used. Since the matcher only returns subsumption relations for classes in the conference track, the system is only used in KG track.

\subsection{Adjustments to the Evaluation}

\begin{figure}[t]
    \centering
    \begin{subfigure}{0.49\linewidth}
         \centering
         \includegraphics[width=0.6\linewidth]{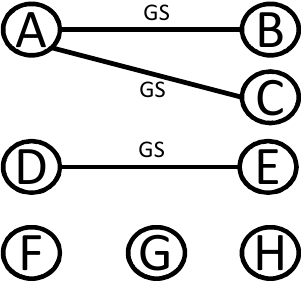}
         \caption{Gold standard: Data sources as nodes and available gold standard (GS) between them as edges.}
         \label{fig:distractors}
    \end{subfigure}
    \hfill
    \begin{subfigure}{0.49\linewidth}
        \centering
        \includegraphics[width=0.7\linewidth]{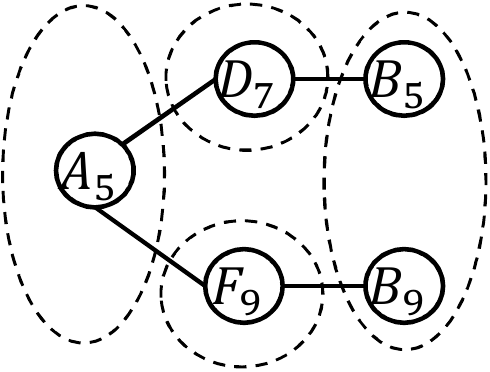}
        \caption{Exemplary system alignment: The nodes represent resources whereas the dashed circles represent the data sources. The edges are correspondences which forms the alignment.}
        \label{fig:alignment_cross}
    \end{subfigure}
    \caption{Adjustments to the evaluation for multi source.}
\end{figure}

All multi source approaches presented in Section \ref{sec:approach} are evaluated in each track and with each binary matching system.
The reference alignment (gold standard) only exists between two data sources (KGs). An example is shown in Figure~\ref{fig:distractors} where only a reference alignment exists for data sources A-B, A-C, and D-E. Thus the evaluation does not analyze the resulting cluster of entities, but the returned alignment which consists of correspondences between two concepts.

Depending on the strategy used, the matching systems may match pairs of KGs for which a gold standard alignment does not exist, or not match pairs for which the gold standard exists. Such an example is shown in Fig.~\ref{fig:alignment_cross}. There is no gold standard for the sources that have been matched (i.e., A-D, A-F, B-D, and B-F), but a gold standard for a pair which has not been matched directly (i.e., A-B).

In order to compare the systems' output to the gold standard, we split the resulting alignment which consists of correspondences between any pair of sources, e.g. $\langle A_5, D_7\rangle$, into individual alignments which correspond to the test case and thus to the gold standard. That splitting is done based on the matched concepts' URIs.
All correspondences which match a concept from A to a concept originating from source B are put into the subset alignment between A and B. With such a setup, the evaluation of a multi source matching system is reduced to a set of pairwise evaluations (therefore the same evaluation techniques can be used). 

At the same time, for strategies that do not perform a pairwise matching of all data sources, we need to compute the transitive closure of all alignments found. In the example, this would include $\langle A_5, B_5\rangle$ and $\langle A_5, B_9\rangle$.
Thus, each multi-source matching strategy can let the evaluation system know if a transitive closure of the alignment is needed or not.
If the system requires a transitive closure, that closure is first computed based on the resulting alignment of the multi source matching system.
The algorithm for computing the closure is based on \cite{transitiveclosure} (Section 3.3). Our implementation does not use a database and keeps all data structures in memory to speed up the computation and also allows to add multiple elements which already belong to the same cluster. 

After the transitive closure of the alignment is calculated, each cluster consists of similar entities (this corresponds to all resources shown in Figure \ref{fig:alignment_cross}).
Those have to be mapped to the corresponding test cases. 
In our example, where the result for the pair A-B is to be computed, all combinations of elements from A and B in the same cluster are considered, i.e., $\langle A_5, B_5, =, 1.0\rangle$ and $\langle A_5, B_9, =, 1.0\rangle$. Due to the fact that the confidence is not taken into account for this evaluation, it is set to $1.0$.

Furthermore, for all tracks, a simple and an extended version exist. For the former, only the sources which participate in the gold standard are used.
In the example shown in Figure~\ref{fig:distractors} this will include only the sources A to E.
For the extended version, additional data sources are added for which no reference alignment exists (these are KGs F, G, and H). 
In the conference track, all 16 ontologies are used, whereas for the KG track, we doubled the number of sources to be also 16. The added KGs are randomly drawn from the set of extracted Wikis available in Fandom which are not too small.

\subsection{Results}

\begin{table*}
	\centering
	\caption{Micro F-measure results for each approach and base matcher in the KG track. The extended version has eight more sources to be matched. Runtime is given has HH:MM:SS. TP means transitive pairs and IM represents the incremental merge approach.}
	\label{tab:KGtrackResults}
	
	\scalebox{0.86}{
	\begin{tabular}{|l|l|l|l|l|l|l|l|l|l|l|l|l|l|l|l|l|}
		
		& \multicolumn{8} {c|} {\bfseries Knowledge Graph} & \multicolumn{8} {c|} {\bfseries Knowledge Graph Extended}\\\hline
		& \multicolumn{2} {c|} {\bfseries ALOD2Vec} & \multicolumn{2} {c|} {\bfseries AML} & \multicolumn{2} {c|} {\bfseries ATBox} & \multicolumn{2} {c|} {\bfseries Paris}
		& \multicolumn{2} {c|} {\bfseries ALOD2Vec} & \multicolumn{2} {c|} {\bfseries AML} & \multicolumn{2} {c|} {\bfseries ATBox} & \multicolumn{2} {c|} {\bfseries Paris} \\

		\hline
		Approach                & $F_1$ & Time     & $F_1$ & Time     & $F_1$ & Time     & $F_1$ & Time  & $F_1$ & Time     & $F_1$ & Time     & $F_1$ & Time     & $F_1$ & Time    \\ \hline
		(i) All Pairs           & \textbf{.89}  & 00:53:23          & \textbf{.87}  & 09:55:07 & \textbf{.89}  & 01:27:29 & \textbf{.90}  & 03:34:50 & \textbf{.89}   & 07:11:30 & \textbf{.87}   & 27:48:08 & \textbf{.89}   & 03:31:33 & \textbf{.90}   & 08:31:02 \\ \hline
		(ii) TP: Windowing      &               &                   &       &          &       &          &       &  &       &          &       &          &       &          &       &          \\		
		~ClassesDescending      & .13           & 00:12:42          & .12  & 02:58:27 & .12  & 00:24:51 & .15  & 00:57:05& .13   & 00:54:30 & .12   & 05:03:45 & .12   & 00:35:44 & .15   & 01:02:06 \\		
		~InstancesDescending    & .69           & 00:20:14          & .67  & 04:29:20 & .71  & 00:19:36 & .70  & 01:11:41& .69   & 01:16:37 & .67   & 06:21:59 & .71   & 00:26:34 & .70   & 01:01:48 \\		
		~ModelSizeDescending    & .70           & 00:23:22          & .68  & 03:46:05 & .72  & 00:21:53 & .70  & 00:50:09& .70   & 00:32:58 & .68   & 07:58:39 & .72   & 00:25:50 & .70   & 01:05:12 \\ 		
		~ClassesAscending       & .13           & 00:12:19          & .12  & 03:24:10 & .12  & 00:25:08 & .15  & 01:00:50& .13   & 00:53:19 & .12   & 06:46:31 & .12   & 00:28:15 & .15   & 00:59:49 \\
		~InstancesAscending     & .69           & 00:11:05          & .67  & 04:32:34 & .71  & 00:21:11 & .70  & 01:11:45& .69   & 01:18:40 & .67   & 06:10:51 & .71   & 00:27:41 & .70   & 01:34:20 \\
		~ModelSizeAscending     & .70           & 01:02:47          & .68  & 04:09:27 & .72  & 00:24:06 & .70  & 01:00:18& .70   & 00:37:11 & .68   & 04:46:51 & .71   & 00:35:50 & .70   & 01:03:05 \\ \hline
		         
		(iii) TP: First vs Rest &               &                   &       &          &       &          &       &   &       &          &       &          &       &          &       &        \\		
		~ClassesDescending      & .37           & 00:15:48          & .34  & 02:15:05 & .37  & 00:24:28 & .48  & 00:44:53& .03   & 00:30:38 & .03   & 04:04:29 & .04   & 00:21:26 & .04   & 00:34:25 \\
		~InstancesDescending    & .26           & 00:32:34          & .23  & 03:46:40 & .23  & 00:40:26 & .31  & 01:13:45& .26   & 00:45:10 & .23   & 03:52:26 & .23   & 01:13:12 & .31   & 02:12:17 \\
		~ModelSizeDescending    & .37           & 00:44:45          & .36  & 03:55:28 & .37  & 00:40:08 & .36  & 01:56:42& .37   & 01:57:41 & .36   & 03:54:41 & .37   & 01:10:15 & .36   & 03:26:57 \\ 
		~ClassesAscending       & .18           & 00:17:51          & .16  & 01:12:21 & .15  & 00:24:29 & .21  & 00:35:11& .00   & 00:20:50 & .01   & 00:53:40 & .01   & 00:30:47 & .00   & 00:36:18 \\
		~InstancesAscending     & .18           & 00:07:23 & .17  & 01:32:15 & .18  & 00:13:26 & .18  & 00:32:36& .00   & 00:22:48 & .00   & 00:34:07 & .01   & 00:18:14 & .08   & 00:30:55 \\
		~ModelSizeAscending     & .18           & 00:06:50 & .17  & 01:34:42 & .18  & 00:15:50 & .18  & 00:31:40& .00   & 00:35:18 & .00   & 00:54:24 & .01   & 00:27:59 & .08   & 00:29:47 \\\hline		
		
		(iv) TP: Similarity     & \textbf{.88}  & 00:21:54          & \textbf{.87}  & 02:23:13 & \textbf{.89}  & 00:42:51 & \textbf{.90}  & 01:41:53& .87   & 02:20:39 & \textbf{.86}   & 04:28:56 & \textbf{.88}   & 00:51:23 & \textbf{.89}   & 02:41:35 \\ \hline
		(v) IM: Order Based     &               &                   &       &          &       &          &       &         &       &          &       &          &       &          &       &     \\
		~ClassesDescending      & .83           & 00:23:31          & .85  & 07:01:14 & .87  & 01:20:43 & .85  & 01:44:37& .83   & 01:07:00 & .85   & 08:43:22 & .87   & 02:54:05 & .85   & 03:44:30 \\
		~InstancesDescending    & .85           & 00:54:14          & .85  & 06:07:40 & \textbf{.88}  & 01:14:02 & .86  & 01:57:58& .85   & 02:48:12 & .85   & 09:45:19 & \textbf{.88}   & 02:40:57 & .86   & 03:27:00 \\
		~ModelSizeDescending    & .85           & 00:55:31          & .85  & 08:44:38 & \textbf{.88}  & 01:25:44 & .86  & 02:00:42& .85   & 03:30:22 & .85   & 10:26:52 & \textbf{.88}   & 03:02:54 & .86   & 03:24:05 \\ 
		~ClassesAscending       & .86           & 00:17:30          & .85  & 04:55:48 & .87  & 00:45:12 & .85  & 01:44:41& .86   & 00:38:14 & .85   & 06:36:17 & .86   & 01:11:08 & .84   & 02:06:36 \\
		~InstancesAscending     & .87           & 00:29:42          & .85  & 06:20:09 & .87  & 00:39:30 & .84  & 01:22:59& .86   & 01:17:55 & .85   & 04:42:02 & .86   & 00:55:52 & .83   & 01:39:59 \\
		~ModelSizeAscending     & .87           & 02:10:36          & .85  & 05:02:27 & .87  & 00:35:36 & .84  & 01:01:32& .86   & 01:07:58 & .84   & 05:50:58 & .86   & 01:06:24 & .83   & 01:53:46 \\
		\hline                    
		(vi) IM: Similarity     &               &                   &       &          &       &          &       &       &       &          &       &          &       &          &       &       \\
		~Single                 & .87           & 00:53:23          & \textbf{.86}  & 06:38:20 & .87  & 01:06:03 & \textbf{.87}  & 01:45:28 & .87   & 03:34:44 & \textbf{.86}   & 06:52:42 & .87   & 02:31:10 & .86   & 03:42:16 \\
		~Average                & \textbf{.88}  & 01:35:48          & \textbf{.86}  & 05:13:00 & .87  & 01:00:53 & \textbf{.87}  & 01:21:07& \textbf{.88}   & 02:34:12 & \textbf{.86}   & 07:59:01 & .87   & 02:03:42 & .87   & 03:47:13 \\
		~Complete               & \textbf{.88}  & 01:03:47          & \textbf{.86}  & 05:13:00 & .87  & 01:00:53 & \textbf{.87}  & 01:21:07& \textbf{.88}   & 02:15:35 & \textbf{.86}   & 05:42:25 & .84   & 01:42:01 & .86   & 03:12:53 \\ \hline
	\end{tabular}
	}
\end{table*}

All detailed results can be analyzed in a CSV file uploaded to Github\footnote{\url{https://github.com/dwslab/melt/tree/master/examples/multisourceExperiment/multisourceResults}}. It further distinguishes the class, property, and instance alignments and also shows the precision, recall, and $F_1$ measure computed with micro and macro averages.
All runtimes are comparable with each other because the time used to compute the merged knowledge graph for incremental merge based approaches is included as well.

\subsubsection{Knowledge Graph Track}

Table \ref{tab:KGtrackResults} shows the results for each approach in the KG track.
The AML matcher is the slowest one and needs over 27 hours for matching all pairs in the extended version.

Thus, other approaches need to be applied. In general, approach (iii) performed the worst. The main reason is that it fixes one source and matches all others against it. Thus, the initially selected source places an upper limit on the set of concepts are able to be matched. Furthermore, the approach is very unstable because it relies on a good choice of a matching hub. It turns out that neither the classes nor the instances are a good measure but the total number of statements in the KG (\texttt{Model\-Size}) return better results. When using the latter measure, approach (iii) can reach 0.37 in terms of $F_1$. 

A better approach is to match different sources together like in (ii) where the order is used in combination with a windowing approach. Again, it is better to use an ordering based on the model size. Note here that descending and ascending results in the same execution plan with the only difference that source and target are interchanged (the same results support the assumption that the matchers do not care about switching source and target). 

When using the textual similarity for generating the order, results are getting even better. This holds for the transitive pairs (iii) as well as incremental merge (vi). The former one has very good results, but one has to be careful because the reference alignments exist for KGs which are also in the same topic. Thus, the incremental approach is better suited because all concepts are kept in the union and can be matched in later steps. The runtime increases in contrast to the transitive pair approaches because the matchers used in the experiments do not have a warm start feature and need to read the inputs over and over again (the merge itself also takes some time but everything is included in the final runtime). The base matchers behave similarly with the exception of ALOD2Vec which has better results in (v) when using the ascending order. 

Overall, we can observe that the decomposition approaches can achieve results which are only one percentage point away from the pairwise approaches, but at significantly lower runtimes.

\begin{table}
	\centering
	\caption{Micro F-measure results for each approach and base matcher in the extended conference track. The runtime is given has HH:MM:SS. TP means transitive pairs and IM represents the incremental merge approach.}
	\label{tab:conferenceTrackResults}
	\scalebox{0.7}{
    \begin{tabular}{|l|l|l|l|l|l|l|l|l|}
		& \multicolumn{2} {c|} {\bfseries ALOD2Vec} & \multicolumn{2} {c|} {\bfseries AML} & \multicolumn{2} {c|} {\bfseries ATBox} & \multicolumn{2} {c|} {\bfseries LogMap}\\
		\hline
		Approach                & $F_1$ & Time     & $F_1$ & Time     & $F_1$ & Time     & $F_1$ & Time     \\ \hline
		(i) All Pairs           & \textbf{.56}   & 00:16:26 & \textbf{.74}   & 00:18:25 & \textbf{.60}   & 00:21:08 & \textbf{.68}   & 00:16:45 \\ \hline
		(ii) TP: Windowing      &       &          &       &          &       &          &       &          \\
		~ClassesDescending      & .24   & 00:02:07 & .26   & 00:02:19 & .28   & 00:02:45 & .25   & 00:01:54 \\
		~ModelSizeDescending    & .31   & 00:02:15 & .36   & 00:02:14 & .31   & 00:02:41 & .34   & 00:02:03 \\
		~ClassesAscending       & .22   & 00:02:13 & .22   & 00:02:14 & .28   & 00:02:46 & .24   & 00:02:06 \\
		~ModelSizeAscending     & .31   & 00:02:13 & .36   & 00:02:14 & .31   & 00:02:46 & .34   & 00:01:53 \\ \hline
		(iii) TP: First vs Rest &       &          &       &          &       &          &       &          \\
		~ClassesDescending      & .41   & 00:02:20 & .46   & 00:02:31 & .46   & 00:02:55 & .47   & 00:02:02 \\
		~ModelSizeDescending    & .45   & 00:02:19 & .57   & 00:02:28 & .52   & 00:03:03 & .57   & 00:02:07 \\
		~ClassesAscending       & .37   & 00:02:11 & .42   & 00:02:07 & .39   & 00:02:32 & .39   & 00:02:05 \\
		~ModelSizeAscending     & .37   & 00:02:09 & .42   & 00:02:05 & .39   & 00:02:33 & .39   & 00:02:05 \\ \hline
		(iv) TP: Similarity     & .39   & 00:02:12 & .47   & 00:02:16 & .41   & 00:02:46 & .43   & 00:02:02 \\ \hline
		(v) IM: Order Based     &       &          &       &          &       &          &       &          \\
		~ClassesDescending      & \textbf{.54}   & 00:02:57 & .64   & 00:04:27 & .57   & 00:06:16 & .63   & 00:02:19 \\
		~ModelSizeDescending    & \textbf{.56}   & 00:02:51 & .65   & 00:04:26 & \textbf{.58}   & 00:06:12 & .63   & 00:02:13 \\
		~ClassesAscending       & .51   & 00:02:43 & .64   & 00:03:21 & .56   & 00:04:48 & .63   & 00:02:13 \\
		~ModelSizeAscending     & .51   & 00:02:40 & \textbf{.66}   & 00:03:22 & .54   & 00:04:43 & .63   & 00:02:12 \\ \hline
		(vi) IM: Similarity     &       &          &       &          &       &          &       &          \\
		Single                  & .48   & 00:02:35 & .60   & 00:04:52 & .52   & 00:05:01 & \textbf{.64}   & 00:02:13 \\
		Average                 & .50   & 00:02:36 & .59   & 00:04:37 & .52   & 00:04:52 & .63   & 00:02:11 \\
		Complete                & .42   & 00:02:31 & .xx   & xx:xx:xx & .50   & 00:03:52 & \textbf{.64}   & 00:02:08 \\ \hline
    \end{tabular}
    }
\end{table}

\subsubsection{Conference Track}

For the conference track, Table \ref{tab:conferenceTrackResults} lists the results for the extended version which contains 16 ontologies.

The results are comparable to the one from the KG track, despite the drop in $F_1$ being higher compared to the all pairs approach.
There are also some smaller differences. For example, the transitive pairs approach where one source is fixed (iii) is consistently better than the windowing approach (ii). This is the case because all sources have a similar topic and fixing one of it as a matching hub is not so severe. But even if the sources all have a similar topic, a clustering of sources based on text makes sense, and LogMap outperforms all other approaches with it except approach (i) all pairs. The execution of AML for approach (vi - complete) was canceled after four hours (which would usually only need about five minutes).

To sum up, the incremental approaches are usually better than the transitive pairs. The model size with the decending order yields better results than any other ordering for most of the matchers.
As an alternative, the textual clustering (vi) is also a good candidate to generate near optimal results.

\subsubsection{Clustering repair approaches}

\begin{table}[t]
	\centering
	\caption{Results for all tracks when applying different clustering repair approaches on the approach all pairs (i). The values represent micro $F_1$ measure.}
	\label{tab:clustering}
    \scalebox{0.89}{
    \begin{tabular}{|l|l|l|l|l|l|l|l|l|}
		& {\bfseries ALOD2Vec} & {\bfseries AML} & {\bfseries ATBox} & {\bfseries LogMap}\\
		\hline
		Conference   &      &      &      &      \\
		~~All Pairs (Baseline)    & .563 & \textbf{.739} & .595 & .683  \\
        ~~CLIP         & \textbf{.580} & .729 & \textbf{.598} & .663  \\
        ~~Error 0.90        & \textbf{.593} & \textbf{.739} & .595 & .683  \\
        ~~Connected Components & .563 & \textbf{.739} & .595 & \textbf{.688}  \\
        \hline
        Conference extended   &      &      &   & \\
        ~~All Pairs (Baseline)         & .563          & \textbf{.739} & .595          & \textbf{.683}  \\ 
        ~~Error 0.90        & \textbf{.577} & \textbf{.739} & \textbf{.609} & .676  \\
        \hline
        & {\bfseries ALOD2Vec} & {\bfseries AML} & {\bfseries ATBox} & {\bfseries Paris}\\
        \hline
        Knowledge graph   &      &      &   & \\
        ~~All Pairs (Baseline)   & \textbf{.890} & .870          & \textbf{.891} & \textbf{.896} \\
        ~~CLIP        & .889          & \textbf{.872} & \textbf{.891} & \textbf{.896} \\\hline
    \end{tabular}
    }
\end{table}

\begin{figure}
	\centering
	\includegraphics[width=0.25\linewidth]{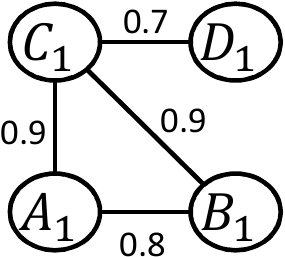}
	\caption{An example for a similarity graph with four concepts from different sources (indicated by the capital letter). $A_1$, $B_1$, and $C_1$ corresponds to one cluster, whereas the correspondence between $C_1$ and $D_1$ is probably wrong.}
	\label{fig:cluster_repair}
\end{figure}

One advantage of the all pairs approach (i) is the computed similarity graph between concepts. It contains the entities as nodes and the edges represent the correspondences created by a matching system.
Due to the fact that all data sources are aligned with each other, the resulting graph can be used to further analyze and remove incorrectly stated correspondences.
An example is given in Figure~\ref{fig:cluster_repair}. It shows that the correspondence between $C_1$ and $D_1$ can be removed because all similar resource $A_1$, $B_1$, and $C_1$ are not matched to $D_1$ even tough all data sources A, B, C, and D are matched together.

In this section, it is analyzed if the all pairs approach (i) followed by such a conflict resolution approach is worth the additional runtime it requires. 
Therefore, we reimplemented the approach from \cite{raad2018detecting}. It computes an error degree for each correspondence and we filter them based on the following thresholds: 0.99, 0.95, 0.90, 0.80, 0.60.
Additionally, all clustering approches from FAMER \cite{famer} are tried out as well to further increase the performance.
They need multiple parameters such as a priority selection. Therefore, we tried out all possible values and report the best ones in Table~\ref{tab:clustering}.
CLIP, and Connected Components refers to FAMER based clustering approaches and Error 0.90 refers to \cite{raad2018detecting} with the corresponding error threshold.

The gains highly depend on the base matcher and track. The performance of ALOD2Vec can be improved by three percentage points in term of $F_1$ for the conference track when removing all correspondence with an error greater than 0.9. For ATBox, the gain is 0.3 when using the CLIP approach and 0.5 for LogMap when using Connected Components.
For the KG track the CLIP approach works better but yields only an improvement of 0.2 $F_1$ points for AML.
All in all, it can be shown that the results cannot be further improved by a large margin except for ALOD2Vec in both conference tracks.

\section{Conclusion and Outlook}
\label{sec:conclusion}

In this paper we showed that the order and the multi-source strategy play an important role when matching multiple KGs.
Two approaches turned out to work in most scenarios: 1) incremental merge with model size descending and 2) incremental merge with textual similarity (average link).
They have the best trade-off between quality (overall F-measure) and efficiency (reduced runtime).
Thus, we propose to first try out these methods to reuse an existing matching system in a multi-source setting. It reduces the quadratic time from the all pairs approach to a linear one with a near-optimal result.
All methods and evaluation capabilities are implemented and documented in the open-source matching framework MELT to allow other researchers to reproduce and reuse the proposed approaches.

\begin{figure}[t]
	\centering
	\includegraphics[width=0.55\linewidth]{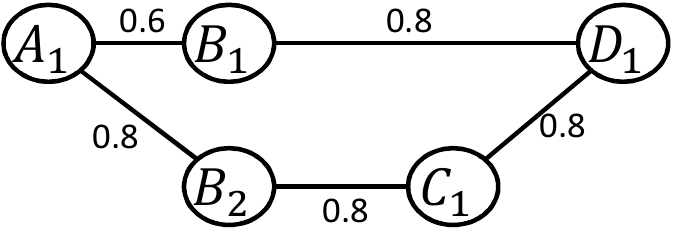}
	\caption{Two paths between entity $A_1$ and $D_1$ which are in the same cluster under the transitive closure. The confidence of the correspondence between $A_1$ and $D_1$ can be computed in multiple ways.}
	\label{fig:correspondence}
\end{figure}

Current matching systems usually do not output only matches, but also confidence scores, which are then used, e.g., for downstream filtering. So far, especially in the transitive pairs approaches, we have not yet analyzed different alternatives for computing such scores.  An example is depicted in Figure \ref{fig:correspondence}. It is a priori not clear what the confidence between $A_1$ and $D_1$ should be. In the current evaluation we just set it to the constant value of 1.0. It is important to note that in our work the correspondences are evaluated regardless of their confidence value and can thus be set to an arbitrary value. Nonetheless it is an interesting direction for researchers which require such values.

When matching a larger scale of input knowledge graphs, parallelization becomes an interesting option. While transitive pairs approaches can be easily parallelized, this is not so straight forward for the incremental merge based approaches. Here, finding a similarity measure which does not only yield good overall results, but also a tree with a low height is essential to leverage the full potential of parallelization.

As a next step, all KGs from the DBkWik project \cite{dbkwik} should be matched together to create one consolidated KG.
But matching is only one part during data integration. Afterwards the sources may provide different conflicting data values.
A first step would be to find out which properties are functional and should only provide one value. For these cases conflict resolution methods should be applied.

\bibliographystyle{ACM-Reference-Format}
\bibliography{multisourcematching}


\begin{thebibliography}{41}


\ifx \showCODEN    \undefined \def \showCODEN     #1{\unskip}     \fi
\ifx \showDOI      \undefined \def \showDOI       #1{#1}\fi
\ifx \showISBNx    \undefined \def \showISBNx     #1{\unskip}     \fi
\ifx \showISBNxiii \undefined \def \showISBNxiii  #1{\unskip}     \fi
\ifx \showISSN     \undefined \def \showISSN      #1{\unskip}     \fi
\ifx \showLCCN     \undefined \def \showLCCN      #1{\unskip}     \fi
\ifx \shownote     \undefined \def \shownote      #1{#1}          \fi
\ifx \showarticletitle \undefined \def \showarticletitle #1{#1}   \fi
\ifx \showURL      \undefined \def \showURL       {\relax}        \fi
\providecommand\bibfield[2]{#2}
\providecommand\bibinfo[2]{#2}
\providecommand\natexlab[1]{#1}
\providecommand\showeprint[2][]{arXiv:#2}

\bibitem[\protect\citeauthoryear{Algergawy, Cheatham, Faria, Ferrara,
  Fundulaki, Harrow, Hertling, Jim{\'e}nez-Ruiz, Karam, Khiat,
  et~al\mbox{.}}{Algergawy et~al\mbox{.}}{2018}]%
        {oaei2018}
\bibfield{author}{\bibinfo{person}{Alsayed Algergawy},
  \bibinfo{person}{Michelle Cheatham}, \bibinfo{person}{Daniel Faria},
  \bibinfo{person}{Alfio Ferrara}, \bibinfo{person}{Irini Fundulaki},
  \bibinfo{person}{Ian Harrow}, \bibinfo{person}{Sven Hertling},
  \bibinfo{person}{Ernesto Jim{\'e}nez-Ruiz}, \bibinfo{person}{Naouel Karam},
  \bibinfo{person}{Abderrahmane Khiat}, {et~al\mbox{.}}}
  \bibinfo{year}{2018}\natexlab{}.
\newblock \showarticletitle{Results of the ontology alignment evaluation
  initiative 2018}. In \bibinfo{booktitle}{\emph{Ontology Matching Workshop @
  ISWC}}, Vol.~\bibinfo{volume}{2288}. \bibinfo{pages}{76--116}.
\newblock


\bibitem[\protect\citeauthoryear{Auer, Bizer, Kobilarov, Lehmann, Cyganiak, and
  Ives}{Auer et~al\mbox{.}}{2007}]%
        {auer2007dbpedia}
\bibfield{author}{\bibinfo{person}{S{\"o}ren Auer}, \bibinfo{person}{Christian
  Bizer}, \bibinfo{person}{Georgi Kobilarov}, \bibinfo{person}{Jens Lehmann},
  \bibinfo{person}{Richard Cyganiak}, {and} \bibinfo{person}{Zachary Ives}.}
  \bibinfo{year}{2007}\natexlab{}.
\newblock \showarticletitle{Dbpedia: A nucleus for a web of open data}.
\newblock In \bibinfo{booktitle}{\emph{The semantic web}}.
  \bibinfo{pages}{722--735}.
\newblock


\bibitem[\protect\citeauthoryear{Beek, Raad, Wielemaker, and Van~Harmelen}{Beek
  et~al\mbox{.}}{2018}]%
        {transitiveclosure}
\bibfield{author}{\bibinfo{person}{Wouter Beek}, \bibinfo{person}{Joe Raad},
  \bibinfo{person}{Jan Wielemaker}, {and} \bibinfo{person}{Frank
  Van~Harmelen}.} \bibinfo{year}{2018}\natexlab{}.
\newblock \showarticletitle{sameas. cc: The closure of 500m owl: sameas
  statements}. In \bibinfo{booktitle}{\emph{ESWC}}. \bibinfo{pages}{65--80}.
\newblock


\bibitem[\protect\citeauthoryear{Bizer, Lehmann, Kobilarov, Auer, Becker,
  Cyganiak, and Hellmann}{Bizer et~al\mbox{.}}{2009}]%
        {dbpedia}
\bibfield{author}{\bibinfo{person}{Christian Bizer}, \bibinfo{person}{Jens
  Lehmann}, \bibinfo{person}{Georgi Kobilarov}, \bibinfo{person}{S{\"o}ren
  Auer}, \bibinfo{person}{Christian Becker}, \bibinfo{person}{Richard
  Cyganiak}, {and} \bibinfo{person}{Sebastian Hellmann}.}
  \bibinfo{year}{2009}\natexlab{}.
\newblock \showarticletitle{DBpedia-A crystallization point for the Web of
  Data}.
\newblock \bibinfo{journal}{\emph{Journal of web semantics}}
  \bibinfo{volume}{7}, \bibinfo{number}{3} (\bibinfo{year}{2009}),
  \bibinfo{pages}{154--165}.
\newblock


\bibitem[\protect\citeauthoryear{Blondel, Guillaume, Lambiotte, and
  Lefebvre}{Blondel et~al\mbox{.}}{2008}]%
        {blondel2008fast}
\bibfield{author}{\bibinfo{person}{Vincent~D Blondel},
  \bibinfo{person}{Jean-Loup Guillaume}, \bibinfo{person}{Renaud Lambiotte},
  {and} \bibinfo{person}{Etienne Lefebvre}.} \bibinfo{year}{2008}\natexlab{}.
\newblock \showarticletitle{Fast unfolding of communities in large networks}.
\newblock \bibinfo{journal}{\emph{Journal of statistical mechanics: theory and
  experiment}} \bibinfo{volume}{2008}, \bibinfo{number}{10}
  (\bibinfo{year}{2008}), \bibinfo{pages}{P10008}.
\newblock


\bibitem[\protect\citeauthoryear{Cheatham and Hitzler}{Cheatham and
  Hitzler}{2014}]%
        {oaeiconference}
\bibfield{author}{\bibinfo{person}{Michelle Cheatham} {and}
  \bibinfo{person}{Pascal Hitzler}.} \bibinfo{year}{2014}\natexlab{}.
\newblock \showarticletitle{Conference v2. 0: An uncertain version of the oaei
  conference benchmark}. In \bibinfo{booktitle}{\emph{ISWC}}.
  \bibinfo{pages}{33--48}.
\newblock


\bibitem[\protect\citeauthoryear{Christophides, Efthymiou, Palpanas, Papadakis,
  and Stefanidis}{Christophides et~al\mbox{.}}{2020}]%
        {survey}
\bibfield{author}{\bibinfo{person}{Vassilis Christophides},
  \bibinfo{person}{Vasilis Efthymiou}, \bibinfo{person}{Themis Palpanas},
  \bibinfo{person}{George Papadakis}, {and} \bibinfo{person}{Kostas
  Stefanidis}.} \bibinfo{year}{2020}\natexlab{}.
\newblock \showarticletitle{An Overview of End-to-End Entity Resolution for Big
  Data}.
\newblock \bibinfo{journal}{\emph{Comput. Surveys}} \bibinfo{volume}{53},
  \bibinfo{number}{6} (\bibinfo{year}{2020}).
\newblock


\bibitem[\protect\citeauthoryear{Dong, Gabrilovich, Heitz, Horn, Lao, Murphy,
  Strohmann, Sun, and Zhang}{Dong et~al\mbox{.}}{2014}]%
        {dong2014knowledge}
\bibfield{author}{\bibinfo{person}{Xin Dong}, \bibinfo{person}{Evgeniy
  Gabrilovich}, \bibinfo{person}{Geremy Heitz}, \bibinfo{person}{Wilko Horn},
  \bibinfo{person}{Ni Lao}, \bibinfo{person}{Kevin Murphy},
  \bibinfo{person}{Thomas Strohmann}, \bibinfo{person}{Shaohua Sun}, {and}
  \bibinfo{person}{Wei Zhang}.} \bibinfo{year}{2014}\natexlab{}.
\newblock \showarticletitle{Knowledge vault: A web-scale approach to
  probabilistic knowledge fusion}. In \bibinfo{booktitle}{\emph{SIGKDD}}.
  \bibinfo{pages}{601--610}.
\newblock


\bibitem[\protect\citeauthoryear{Elmagarmid, Ipeirotis, and
  Verykios}{Elmagarmid et~al\mbox{.}}{2006}]%
        {elmagarmid2006duplicate}
\bibfield{author}{\bibinfo{person}{Ahmed~K Elmagarmid},
  \bibinfo{person}{Panagiotis~G Ipeirotis}, {and} \bibinfo{person}{Vassilios~S
  Verykios}.} \bibinfo{year}{2006}\natexlab{}.
\newblock \showarticletitle{Duplicate record detection: A survey}.
\newblock \bibinfo{journal}{\emph{IEEE Transactions on knowledge and data
  engineering}} \bibinfo{volume}{19}, \bibinfo{number}{1}
  (\bibinfo{year}{2006}), \bibinfo{pages}{1--16}.
\newblock


\bibitem[\protect\citeauthoryear{Eppstein}{Eppstein}{2000}]%
        {hierarchicalClustering}
\bibfield{author}{\bibinfo{person}{David Eppstein}.}
  \bibinfo{year}{2000}\natexlab{}.
\newblock \showarticletitle{Fast hierarchical clustering and other applications
  of dynamic closest pairs}.
\newblock \bibinfo{journal}{\emph{Journal of Experimental Algorithmics (JEA)}}
  \bibinfo{volume}{5} (\bibinfo{year}{2000}), \bibinfo{pages}{1--es}.
\newblock


\bibitem[\protect\citeauthoryear{Euzenat, Meilicke, Stuckenschmidt, Shvaiko,
  and Trojahn}{Euzenat et~al\mbox{.}}{2011}]%
        {euzenat2011ontology}
\bibfield{author}{\bibinfo{person}{J{\'e}r{\^o}me Euzenat},
  \bibinfo{person}{Christian Meilicke}, \bibinfo{person}{Heiner
  Stuckenschmidt}, \bibinfo{person}{Pavel Shvaiko}, {and}
  \bibinfo{person}{C{\'a}ssia Trojahn}.} \bibinfo{year}{2011}\natexlab{}.
\newblock \showarticletitle{Ontology alignment evaluation initiative: six years
  of experience}.
\newblock In \bibinfo{booktitle}{\emph{Journal on data semantics XV}}.
  \bibinfo{pages}{158--192}.
\newblock


\bibitem[\protect\citeauthoryear{Euzenat, Shvaiko, et~al\mbox{.}}{Euzenat
  et~al\mbox{.}}{2007}]%
        {euzenat2007}
\bibfield{author}{\bibinfo{person}{J{\'e}r{\^o}me Euzenat},
  \bibinfo{person}{Pavel Shvaiko}, {et~al\mbox{.}}}
  \bibinfo{year}{2007}\natexlab{}.
\newblock \bibinfo{booktitle}{\emph{Ontology matching}}.
  Vol.~\bibinfo{volume}{18}.
\newblock


\bibitem[\protect\citeauthoryear{Fern{\'a}ndez, Beek, Mart{\'\i}nez-Prieto, and
  Arias}{Fern{\'a}ndez et~al\mbox{.}}{2017}]%
        {lodAlot}
\bibfield{author}{\bibinfo{person}{Javier~D Fern{\'a}ndez},
  \bibinfo{person}{Wouter Beek}, \bibinfo{person}{Miguel~A
  Mart{\'\i}nez-Prieto}, {and} \bibinfo{person}{Mario Arias}.}
  \bibinfo{year}{2017}\natexlab{}.
\newblock \showarticletitle{LOD-a-lot}. In \bibinfo{booktitle}{\emph{ISWC}}.
  \bibinfo{pages}{75--83}.
\newblock


\bibitem[\protect\citeauthoryear{Fernández, Martínez-Prieto, Gutiérrez,
  Polleres, and Arias}{Fernández et~al\mbox{.}}{2013}]%
        {HDT}
\bibfield{author}{\bibinfo{person}{Javier~D. Fernández},
  \bibinfo{person}{Miguel~A. Martínez-Prieto}, \bibinfo{person}{Claudio
  Gutiérrez}, \bibinfo{person}{Axel Polleres}, {and} \bibinfo{person}{Mario
  Arias}.} \bibinfo{year}{2013}\natexlab{}.
\newblock \showarticletitle{Binary RDF Representation for Publication and
  Exchange (HDT)}.
\newblock \bibinfo{journal}{\emph{Web Semantics: Science, Services and Agents
  on the World Wide Web}}  \bibinfo{volume}{19} (\bibinfo{year}{2013}),
  \bibinfo{pages}{22–41}.
\newblock


\bibitem[\protect\citeauthoryear{Haas, Hentschel, Kossmann, and Miller}{Haas
  et~al\mbox{.}}{2009}]%
        {haas2009schema}
\bibfield{author}{\bibinfo{person}{Laura~M Haas}, \bibinfo{person}{Martin
  Hentschel}, \bibinfo{person}{Donald Kossmann}, {and}
  \bibinfo{person}{Ren{\'e}e~J Miller}.} \bibinfo{year}{2009}\natexlab{}.
\newblock \showarticletitle{Schema and data: A holistic approach to mapping,
  resolution and fusion in information integration}. In
  \bibinfo{booktitle}{\emph{International Conference on Conceptual Modeling}}.
  \bibinfo{pages}{27--40}.
\newblock


\bibitem[\protect\citeauthoryear{Heist, Hertling, Ringler, and Paulheim}{Heist
  et~al\mbox{.}}{2020}]%
        {heist2020knowledge}
\bibfield{author}{\bibinfo{person}{Nicolas Heist}, \bibinfo{person}{Sven
  Hertling}, \bibinfo{person}{Daniel Ringler}, {and} \bibinfo{person}{Heiko
  Paulheim}.} \bibinfo{year}{2020}\natexlab{}.
\newblock \bibinfo{title}{Knowledge Graphs on the Web-An Overview.}
\newblock
\newblock


\bibitem[\protect\citeauthoryear{Hertling and Paulheim}{Hertling and
  Paulheim}{2018}]%
        {dbkwik}
\bibfield{author}{\bibinfo{person}{Sven Hertling} {and} \bibinfo{person}{Heiko
  Paulheim}.} \bibinfo{year}{2018}\natexlab{}.
\newblock \showarticletitle{Dbkwik: A consolidated knowledge graph from
  thousands of wikis}. In \bibinfo{booktitle}{\emph{2018 IEEE International
  Conference on Big Knowledge (ICBK)}}. \bibinfo{pages}{17--24}.
\newblock


\bibitem[\protect\citeauthoryear{Hertling and Paulheim}{Hertling and
  Paulheim}{2020a}]%
        {atbox2020}
\bibfield{author}{\bibinfo{person}{Sven Hertling} {and} \bibinfo{person}{Heiko
  Paulheim}.} \bibinfo{year}{2020}\natexlab{a}.
\newblock \showarticletitle{ATBox results for OAEI 2020}. In
  \bibinfo{booktitle}{\emph{Ontology Matching Workshop @ ISWC}},
  Vol.~\bibinfo{volume}{2788}. \bibinfo{pages}{168--175}.
\newblock


\bibitem[\protect\citeauthoryear{Hertling and Paulheim}{Hertling and
  Paulheim}{2020b}]%
        {dbkwikResults}
\bibfield{author}{\bibinfo{person}{Sven Hertling} {and} \bibinfo{person}{Heiko
  Paulheim}.} \bibinfo{year}{2020}\natexlab{b}.
\newblock \showarticletitle{The knowledge graph track at OAEI : Gold standards,
  baselines, and the golden hammer bias}. In \bibinfo{booktitle}{\emph{ESWC}},
  Vol.~\bibinfo{volume}{12123}. \bibinfo{pages}{343--359}.
\newblock


\bibitem[\protect\citeauthoryear{Jiménez-Ruiz}{Jiménez-Ruiz}{2020}]%
        {logmap2020}
\bibfield{author}{\bibinfo{person}{Ernesto Jiménez-Ruiz}.}
  \bibinfo{year}{2020}\natexlab{}.
\newblock \showarticletitle{LogMap family participation in the OAEI 2020}. In
  \bibinfo{booktitle}{\emph{Ontology Matching Workshop @ ISWC}},
  Vol.~\bibinfo{volume}{2788}. \bibinfo{pages}{201--203}.
\newblock


\bibitem[\protect\citeauthoryear{Kolb, Thor, and Rahm}{Kolb
  et~al\mbox{.}}{2012}]%
        {kolb2012dedoop}
\bibfield{author}{\bibinfo{person}{Lars Kolb}, \bibinfo{person}{Andreas Thor},
  {and} \bibinfo{person}{Erhard Rahm}.} \bibinfo{year}{2012}\natexlab{}.
\newblock \showarticletitle{Dedoop: Efficient deduplication with hadoop}.
\newblock \bibinfo{journal}{\emph{VLDB}} \bibinfo{volume}{5},
  \bibinfo{number}{12} (\bibinfo{year}{2012}), \bibinfo{pages}{1878--1881}.
\newblock


\bibitem[\protect\citeauthoryear{Kruskal}{Kruskal}{1956}]%
        {kruskal}
\bibfield{author}{\bibinfo{person}{Joseph~B Kruskal}.}
  \bibinfo{year}{1956}\natexlab{}.
\newblock \showarticletitle{On the shortest spanning subtree of a graph and the
  traveling salesman problem}.
\newblock \bibinfo{journal}{\emph{Proceedings of the American Mathematical
  society}} \bibinfo{volume}{7}, \bibinfo{number}{1} (\bibinfo{year}{1956}),
  \bibinfo{pages}{48--50}.
\newblock


\bibitem[\protect\citeauthoryear{Li, Liu, Zhang, Wang, and Wan}{Li
  et~al\mbox{.}}{2020}]%
        {blockingsurvey}
\bibfield{author}{\bibinfo{person}{Bo-Han Li}, \bibinfo{person}{Yi Liu},
  \bibinfo{person}{An-Man Zhang}, \bibinfo{person}{Wen-Huan Wang}, {and}
  \bibinfo{person}{Shuo Wan}.} \bibinfo{year}{2020}\natexlab{}.
\newblock \showarticletitle{A survey on blocking technology of entity
  resolution}.
\newblock \bibinfo{journal}{\emph{Journal of Computer Science and Technology}}
  \bibinfo{volume}{35}, \bibinfo{number}{4} (\bibinfo{year}{2020}),
  \bibinfo{pages}{769--793}.
\newblock


\bibitem[\protect\citeauthoryear{Lima, Faria, Couto, Cruz, and Pesquita}{Lima
  et~al\mbox{.}}{2020}]%
        {aml2020}
\bibfield{author}{\bibinfo{person}{Beatriz Lima}, \bibinfo{person}{Daniel
  Faria}, \bibinfo{person}{Francisco~M Couto}, \bibinfo{person}{Isabel~F Cruz},
  {and} \bibinfo{person}{Catia Pesquita}.} \bibinfo{year}{2020}\natexlab{}.
\newblock \showarticletitle{OAEI 2020 results for AML and AMLC}. In
  \bibinfo{booktitle}{\emph{Ontology Matching Workshop @ ISWC}},
  Vol.~\bibinfo{volume}{2788}. \bibinfo{pages}{154--160}.
\newblock


\bibitem[\protect\citeauthoryear{Nentwig, Gro{\ss}, M{\"o}ller, and
  Rahm}{Nentwig et~al\mbox{.}}{2017a}]%
        {nentwig2017distributed}
\bibfield{author}{\bibinfo{person}{Markus Nentwig}, \bibinfo{person}{Anika
  Gro{\ss}}, \bibinfo{person}{Maximilian M{\"o}ller}, {and}
  \bibinfo{person}{Erhard Rahm}.} \bibinfo{year}{2017}\natexlab{a}.
\newblock \showarticletitle{Distributed holistic clustering on linked data}. In
  \bibinfo{booktitle}{\emph{OTM Confederated International Conferences}}.
  \bibinfo{pages}{371--382}.
\newblock


\bibitem[\protect\citeauthoryear{Nentwig, Hartung, Ngonga~Ngomo, and
  Rahm}{Nentwig et~al\mbox{.}}{2017b}]%
        {linkdiscoverysurvey}
\bibfield{author}{\bibinfo{person}{Markus Nentwig}, \bibinfo{person}{Michael
  Hartung}, \bibinfo{person}{Axel-Cyrille Ngonga~Ngomo}, {and}
  \bibinfo{person}{Erhard Rahm}.} \bibinfo{year}{2017}\natexlab{b}.
\newblock \showarticletitle{A survey of current link discovery frameworks}.
\newblock \bibinfo{journal}{\emph{Semantic Web}} \bibinfo{volume}{8},
  \bibinfo{number}{3} (\bibinfo{year}{2017}), \bibinfo{pages}{419--436}.
\newblock


\bibitem[\protect\citeauthoryear{Paulheim}{Paulheim}{2017}]%
        {KGRefine}
\bibfield{author}{\bibinfo{person}{Heiko Paulheim}.}
  \bibinfo{year}{2017}\natexlab{}.
\newblock \showarticletitle{Knowledge graph refinement: A survey of approaches
  and evaluation methods}.
\newblock \bibinfo{journal}{\emph{Semantic web}} \bibinfo{volume}{8},
  \bibinfo{number}{3} (\bibinfo{year}{2017}), \bibinfo{pages}{489--508}.
\newblock


\bibitem[\protect\citeauthoryear{Portisch, Hladik, and Paulheim}{Portisch
  et~al\mbox{.}}{2020}]%
        {alod2vec2020}
\bibfield{author}{\bibinfo{person}{Jan Portisch}, \bibinfo{person}{Michael
  Hladik}, {and} \bibinfo{person}{Heiko Paulheim}.}
  \bibinfo{year}{2020}\natexlab{}.
\newblock \showarticletitle{ALOD2Vec matcher results for OAEI 2020}. In
  \bibinfo{booktitle}{\emph{Ontology Matching Workshop @ ISWC}},
  Vol.~\bibinfo{volume}{2788}. \bibinfo{pages}{147--153}.
\newblock


\bibitem[\protect\citeauthoryear{Pour, Algergawy, Amini, Faria, Fundulaki,
  Harrow, Hertling, Jim{\'e}nez-Ruiz, Jonquet, Karam, et~al\mbox{.}}{Pour
  et~al\mbox{.}}{2020}]%
        {oaei2020}
\bibfield{author}{\bibinfo{person}{Nikooie Pour}, \bibinfo{person}{Alsayed
  Algergawy}, \bibinfo{person}{Reihaneh Amini}, \bibinfo{person}{Daniel Faria},
  \bibinfo{person}{Irini Fundulaki}, \bibinfo{person}{Ian Harrow},
  \bibinfo{person}{Sven Hertling}, \bibinfo{person}{Ernesto Jim{\'e}nez-Ruiz},
  \bibinfo{person}{Clement Jonquet}, \bibinfo{person}{Naouel Karam},
  {et~al\mbox{.}}} \bibinfo{year}{2020}\natexlab{}.
\newblock \showarticletitle{Results of the ontology alignment evaluation
  initiative 2020}. In \bibinfo{booktitle}{\emph{Ontology Matching Workshop @
  ISWC}}, Vol.~\bibinfo{volume}{2788}. \bibinfo{pages}{92--138}.
\newblock


\bibitem[\protect\citeauthoryear{Raad, Beek, Van~Harmelen, Pernelle, and
  Sa{\"\i}s}{Raad et~al\mbox{.}}{2018}]%
        {raad2018detecting}
\bibfield{author}{\bibinfo{person}{Joe Raad}, \bibinfo{person}{Wouter Beek},
  \bibinfo{person}{Frank Van~Harmelen}, \bibinfo{person}{Nathalie Pernelle},
  {and} \bibinfo{person}{Fatiha Sa{\"\i}s}.} \bibinfo{year}{2018}\natexlab{}.
\newblock \showarticletitle{Detecting erroneous identity links on the web using
  network metrics}. In \bibinfo{booktitle}{\emph{ISWC}}.
  \bibinfo{pages}{391--407}.
\newblock


\bibitem[\protect\citeauthoryear{Rahm}{Rahm}{2016}]%
        {surveyrahm}
\bibfield{author}{\bibinfo{person}{Erhard Rahm}.}
  \bibinfo{year}{2016}\natexlab{}.
\newblock \showarticletitle{The case for holistic data integration}. In
  \bibinfo{booktitle}{\emph{East European Conference on Advances in Databases
  and Information Systems}}. \bibinfo{pages}{11--27}.
\newblock


\bibitem[\protect\citeauthoryear{Saeedi, Peukert, and Rahm}{Saeedi
  et~al\mbox{.}}{2018}]%
        {famer}
\bibfield{author}{\bibinfo{person}{Alieh Saeedi}, \bibinfo{person}{Eric
  Peukert}, {and} \bibinfo{person}{Erhard Rahm}.}
  \bibinfo{year}{2018}\natexlab{}.
\newblock \showarticletitle{Using link features for entity clustering in
  knowledge graphs}. In \bibinfo{booktitle}{\emph{ESWC}}.
  \bibinfo{pages}{576--592}.
\newblock


\bibitem[\protect\citeauthoryear{Schmachtenberg, Bizer, and
  Paulheim}{Schmachtenberg et~al\mbox{.}}{2014}]%
        {schmachtenberg2014adoption}
\bibfield{author}{\bibinfo{person}{Max Schmachtenberg},
  \bibinfo{person}{Christian Bizer}, {and} \bibinfo{person}{Heiko Paulheim}.}
  \bibinfo{year}{2014}\natexlab{}.
\newblock \showarticletitle{Adoption of the linked data best practices in
  different topical domains}. In \bibinfo{booktitle}{\emph{ISWC}}.
  \bibinfo{pages}{245--260}.
\newblock


\bibitem[\protect\citeauthoryear{Stavropoulos, Andreadis, Kontopoulos, and
  Kompatsiaris}{Stavropoulos et~al\mbox{.}}{2019}]%
        {topicDrift}
\bibfield{author}{\bibinfo{person}{Thanos~G Stavropoulos},
  \bibinfo{person}{Stelios Andreadis}, \bibinfo{person}{Efstratios
  Kontopoulos}, {and} \bibinfo{person}{Ioannis Kompatsiaris}.}
  \bibinfo{year}{2019}\natexlab{}.
\newblock \showarticletitle{SemaDrift: A hybrid method and visual tools to
  measure semantic drift in ontologies}.
\newblock \bibinfo{journal}{\emph{Journal of Web Semantics}}
  \bibinfo{volume}{54} (\bibinfo{year}{2019}), \bibinfo{pages}{87--106}.
\newblock


\bibitem[\protect\citeauthoryear{Suchanek, Abiteboul, and Senellart}{Suchanek
  et~al\mbox{.}}{2011}]%
        {suchanek2011paris}
\bibfield{author}{\bibinfo{person}{Fabian~M. Suchanek}, \bibinfo{person}{Serge
  Abiteboul}, {and} \bibinfo{person}{Pierre Senellart}.}
  \bibinfo{year}{2011}\natexlab{}.
\newblock \showarticletitle{{PARIS:} Probabilistic Alignment of Relations,
  Instances, and Schema}.
\newblock \bibinfo{journal}{\emph{VLDB}} \bibinfo{volume}{5},
  \bibinfo{number}{3} (\bibinfo{year}{2011}), \bibinfo{pages}{157--168}.
\newblock


\bibitem[\protect\citeauthoryear{Tanon, Weikum, and Suchanek}{Tanon
  et~al\mbox{.}}{2020}]%
        {yago}
\bibfield{author}{\bibinfo{person}{Thomas~Pellissier Tanon},
  \bibinfo{person}{Gerhard Weikum}, {and} \bibinfo{person}{Fabian Suchanek}.}
  \bibinfo{year}{2020}\natexlab{}.
\newblock \showarticletitle{Yago 4: A reason-able knowledge base}. In
  \bibinfo{booktitle}{\emph{ESWC}}. \bibinfo{pages}{583--596}.
\newblock


\bibitem[\protect\citeauthoryear{Vrande{\v{c}}i{\'c} and
  Kr{\"o}tzsch}{Vrande{\v{c}}i{\'c} and Kr{\"o}tzsch}{2014}]%
        {vrandevcic2014wikidata}
\bibfield{author}{\bibinfo{person}{Denny Vrande{\v{c}}i{\'c}} {and}
  \bibinfo{person}{Markus Kr{\"o}tzsch}.} \bibinfo{year}{2014}\natexlab{}.
\newblock \showarticletitle{Wikidata: a free collaborative knowledgebase}.
\newblock \bibinfo{journal}{\emph{Commun. ACM}} \bibinfo{volume}{57},
  \bibinfo{number}{10} (\bibinfo{year}{2014}), \bibinfo{pages}{78--85}.
\newblock


\bibitem[\protect\citeauthoryear{Wang, Li, Yu, and Feng}{Wang
  et~al\mbox{.}}{2011}]%
        {wang2011entity}
\bibfield{author}{\bibinfo{person}{Jiannan Wang}, \bibinfo{person}{Guoliang
  Li}, \bibinfo{person}{Jeffrey~Xu Yu}, {and} \bibinfo{person}{Jianhua Feng}.}
  \bibinfo{year}{2011}\natexlab{}.
\newblock \showarticletitle{Entity matching: How similar is similar}.
\newblock \bibinfo{journal}{\emph{VLDB}} \bibinfo{volume}{4},
  \bibinfo{number}{10} (\bibinfo{year}{2011}), \bibinfo{pages}{622--633}.
\newblock


\bibitem[\protect\citeauthoryear{Wang, Schlobach, Takens, and
  Van~Atteveldt}{Wang et~al\mbox{.}}{2009}]%
        {topicDriftTwo}
\bibfield{author}{\bibinfo{person}{Shenghui Wang}, \bibinfo{person}{Stefan
  Schlobach}, \bibinfo{person}{Janet Takens}, {and} \bibinfo{person}{Wouter
  Van~Atteveldt}.} \bibinfo{year}{2009}\natexlab{}.
\newblock \showarticletitle{Mapping-chains for studying concept shift in
  political ontologies}.
\newblock \bibinfo{journal}{\emph{Ontology Matching}}  \bibinfo{volume}{13}
  (\bibinfo{year}{2009}).
\newblock


\bibitem[\protect\citeauthoryear{Whang, Menestrina, Koutrika, Theobald, and
  Garcia-Molina}{Whang et~al\mbox{.}}{2009}]%
        {blocking}
\bibfield{author}{\bibinfo{person}{Steven~Euijong Whang},
  \bibinfo{person}{David Menestrina}, \bibinfo{person}{Georgia Koutrika},
  \bibinfo{person}{Martin Theobald}, {and} \bibinfo{person}{Hector
  Garcia-Molina}.} \bibinfo{year}{2009}\natexlab{}.
\newblock \showarticletitle{Entity resolution with iterative blocking}. In
  \bibinfo{booktitle}{\emph{SIGMOD}}. \bibinfo{pages}{219--232}.
\newblock


\bibitem[\protect\citeauthoryear{Zamazal and Sv{\'a}tek}{Zamazal and
  Sv{\'a}tek}{2017}]%
        {conferenceDataset}
\bibfield{author}{\bibinfo{person}{Ond{\v{r}}ej Zamazal} {and}
  \bibinfo{person}{Vojt{\v{e}}ch Sv{\'a}tek}.} \bibinfo{year}{2017}\natexlab{}.
\newblock \showarticletitle{The ten-year ontofarm and its fertilization within
  the onto-sphere}.
\newblock \bibinfo{journal}{\emph{Journal of Web Semantics}}
  \bibinfo{volume}{43} (\bibinfo{year}{2017}), \bibinfo{pages}{46--53}.
\newblock


\end{thebibliography}

\end{document}